\documentclass[pdftex,12pt]{article}
\usepackage{amsmath,amsfonts,amsbsy,latexsym,amssymb,amscd,amstext}
\usepackage{pst-node}
\usepackage{dsfont}
\usepackage{cool}
\usepackage{overpic,youngtab}
\pdfoutput=1
\usepackage{rotating}
\usepackage{tikz}
\usepackage{slashed}

\def\bpm{\begin{pmatrix}}
\def\epm{\end{pmatrix}}
\def\be{\begin{equation}}
\def\cO{\cal O}
\def\ee{\end{equation}}
\def\bea{\begin{eqnarray}}
\def\eea{\end{eqnarray}}
\def\pd{\partial}
\def\a{\alpha}

\def\b{\beta}

\def\d{\delta}
\def\m{\mu}

\def\n{\nu}
\def\t{\tau}

\def\l{\lambda}

\def\r{\rho}

\def\bR{\bar{R}}

\def\bg{\bar{g}}

\def\s{\sigma}
\def\e{\epsilon}

\def\bma{\begin{pmatrix}}
\def\ema{\end{pmatrix}}

\def\bg{\bar{g}}
\def\bi{\begin{itemize}}
\def\ei{\end{itemize}}
\def\bn{\bar{\nabla}}

\def\tr{{\rm tr\,}}

\begin{document}

		\vspace*{-1cm}
		\phantom{hep-ph/***} 
		{\flushleft
			{{FTUAM-22-X}}
			\hfill{{ IFT-UAM/CSIC-22-8}}}
		\vskip 1.5cm
		\begin{center}
		{\LARGE\bfseries   Divergences of the scalar  sector of quadratic gravity.}\\[3mm]
			\vskip .3cm
		
		\end{center}

		\vskip 0.5  cm
		\begin{center}
			{\large Enrique \'Alvarez and Jes\'us Anero.}
			\\
			\vskip .7cm
			{
				Departamento de F\'isica Te\'orica and Instituto de F\'{\i}sica Te\'orica, 
				IFT-UAM/CSIC,\\
				Universidad Aut\'onoma de Madrid, Cantoblanco, 28049, Madrid, Spain\\
				\vskip .1cm

				\vskip .5cm
				
				\begin{minipage}[l]{.9\textwidth}
					\begin{center} 
							\textit{E-mail:} 
						\tt{enrique.alvarez@uam.es},
						\tt{jesusanero@gmail.com}
					\end{center}
				\end{minipage}
			}
		\end{center}
	\thispagestyle{empty}
	
\begin{abstract}
	\noindent
 The divergences coming from a particular sector of gravitational fluctuations around a generic background in general theories of quadratic gravity are analyzed. They can be summarized in a particular type of scalar model, whose properties are analyzed.
\end{abstract}

\newpage
\tableofcontents
	\thispagestyle{empty}
\flushbottom

\newpage
 \section{Introduction. }
In this paper we are going to analyze a particular sector of quantum graviton fluctuations around an arbitrary background when the action is given by general theories  quadratic in spacetime curvature. 
In the particular case where the background metric coincides with the Minkowski one $\bg_{\m\n}= \eta_{\m\n}$,  the decomposition of the graviton fluctuation in terms of spin components  can be expressed in terms of the good old Barnes-Rivers projectors \cite{Barnes}\cite{Alvarez:2018lrg}. Let us briefly recall the formalism. 

Let us define two projectors in momentum space 
\renewcommand{\arraystretch}{1}
\bea
&P^{(0)\b}_\a={k_\a k^\b\over k^2}\equiv \omega_\a\,^\b=\begin{pmatrix}1&0&0&0\\0&0&0&0\\0&0&0&0\\0&0&0&0\end{pmatrix}\nonumber\\
& P^{(1)\b}_\a= \d_\a^\b-{k_\a k^\b\over k^2}\equiv \theta_\a\,^\b=\begin{pmatrix}0&0&0&0\\0&1&0&0\\0&0&1&0\\0&0&0&1\end{pmatrix}
\eea
It should be pointed out  that these operators are non-local in position space where
$\dfrac{1}{k^2}$ stands for $\Box^{-1}$.

As it is well-known, the metric $h_{\m\n}$  transforms in the euclidean setting under the representation $\underline{10}\equiv \yng(2)$ of SO(4), so the spin content and corresponding projectors are given by
\bea
&&\text{s=2}:\quad  h^T_{ij}\equiv h_{ij}-{1\over 3} h\d_{ij}\quad \quad \left(P_2\right)_{\m\n}^{\r\s}\equiv {1\over 2}\left(\theta_\m^\r\theta_\n^\s+\theta_\m^\s\theta_\n^\r\right)-{1\over 3}\theta_{\m\n}\theta^{\r\s}\nonumber\\
&&\text{s=1}:\quad  h_{0i}\nonumber \quad \left(P_1\right)_{\m\n}^{\r\s}\equiv{1\over 2}\left(\theta_\m^\r\omega_\n^\s+\theta_\m^\s\omega_\n^\r+\theta_\n^\r\omega_\m^\s+\theta_\n^\s\omega_\m^\r\right)\\
&&\text{s=0}:\quad  h_{00}\quad \left(P_0^w\right)_{\m\n}^{\r\s}\equiv \omega_{\m\n}\omega^{\r\s}\nonumber\\
&&\text{s=0}:\quad  h\equiv \d^{ij} h_{ij}\quad \left(P_0^s\right)_{\m\n}^{\r\s}\equiv {1\over 3}\theta_{\m\n}\theta^{\r\s}
\eea
these particular projectors  are complete in the symmetrized direct product
\be
Sym\left(T_x\otimes T_x\right)
\ee
where $T_x$ is the tangent space at the point $x\in M$ of the space-time manifold.
\par
It is convenient to define another projector
\be
P_0\equiv P_0^w+P_0^s
\ee
and the non-differential projectors are
\bea
&&I_{\m\n}^{\r\s}\equiv{1\over 2}\left(\d_\m^\r\d_\n^\s+\d_\m^\s \d_\n^\r\right)\nonumber\\
&&T_{\m\n}^{\r\s}\equiv {1\over 4} \eta_{\m\n} \eta^{\r\s}
\eea
then we can write a closure relation for these projectors. To be specific,
\be
\left(P_2\right)_{\m\n}^{\r\s}+\left(P_1\right)_{\m\n}^{\r\s}+\left(P_0\right)_{\m\n}^{\r\s}=I_{\m\n}^{\r\s}
\ee

These projectors are not enough  for some tasks, because  they do not form a basis of the space of four-index tensors of the type of interest. We then need to add a new independent operator
\be
\left(P_0^\times\right)_{\m\n}^{\r\s}={1\over \sqrt{3}}\left(\omega_{\m\n}\theta^{\r\s}+\theta_{\m\n} \omega^{\r\s}\right)\label{PX}
\ee
that can be identified with the mixing of the two spin 0 components, $h$ and $h_{00}$.
It is clear that this new operator cannot be orthogonal to the other four, since closure implies that the only operator orthogonal to the set that closes is the null operator.
\par

This decomposition has been generalized for arbitrary background metric by York \cite{York} by means of his conformally invariant orthogonal decomposition of the graviton fluctuation which reads
\be
h_{\m\n}\equiv h_{\m\n}^{TT}+h^L_{\m\n}+h_{\m\n}^{tr}
\ee
where the longitudinal component is given by
\be
h^L_{\m\n}\equiv \bn_\m \xi_\n+\bn_\n\xi_\m-{2 \over n}\bg_{\m\n}\,\nabla^\l\xi_\l\equiv (L\xi)_{\m\n}
\ee
and 
\be
h^{tr}_{\m\n}\equiv {1\over n} h \bg_{\m\n}
\ee
Besides, in \cite{York} it is demonstrated how to uniquely determine $\xi^\m$ from the knowledge of $h_{\m\n}$.  The transversality condition
\be \nabla^\m  h_{\m\n}^{TT}=0
\ee
leads a covariant equations for the vector field $\xi^\m$
\be
D\xi^\m=-\nabla_\n(L\xi)^{\m\n}
\ee
defining $\Lambda$, the projector into the traceless piece of the graviton fluctuation, this defines a partial differential equation for $\xi$, namely
\be
D\xi=-\bn\cdot \Lambda h
\ee
which uniquely determines $\xi$ in terms of $h_{\m\n}$ unless there are in the background metric conformally Killing vectors, for which $L\xi=0$.
\par
The different spin contributions are mutually orthogonal. It is clear that through a gauge transformation the spin one piece can be completely eliminated (although this is not usually the most convenient way of fixing the gauge).
For example, the usual harmonic (de Donder) background field gauge 
\be
\bn_\m h^{\m\n}={1\over 2} \bn^\n h
\ee
reads
\be
\Box \xi^\m+\bR^{\m\l} \xi_\l=0
\ee
in case the spin one part is eliminated, the residual term proportional to $\bn \xi$ can be absorbed in the spin 0 scalar $\phi$.
\par
This vector can in turn be decomposed as
\be
\xi=\xi_T+\xi_L
\ee
(where $\bn.\xi_T=\bn.\xi=0$).
The {\em scalar component} of the graviton fluctuation comprises  then two parts: $h$ and $\bn.\xi_L$. In this paper we are going to consider the full scalar piece denoted by $\phi\bg_{\m\n}$ just by simplicity. 
\par
Previous related works are to be found in \cite{Stelle} as well as in \cite{Buchbinder}. The renormalizability of the model has been exploited in \cite{Einhorn} to build particle physics models. For its relationship with more general setup of quantum gravity cf. \cite{Windows}. Related scalar models have been studied in \cite{Grinstein}. The existence of runaway solutions in higher order theories has been first pointed out by Dirac \cite{Dirac} in his classic paper, and in the context of quadratic gravity, in \cite{Hawking} \cite{Alvarez:2018lrg}. In this paper we shall use the usual second  order formalism all the way; some points to the much more general first order approach are to be found in \cite{AASG}. Finally, some particular models have enhanced $Diff\times Weyl$ symmetry \cite{Alvarez:2016}. 
\section{Quadratic gravity}
The general lagrangian quadratic in curvature is a combination of monomials in $R^2$, $R_{\m\n}^2$ and $R_{\m\n\r\s}^2$. The most general diff invariant low energy action is then
\be
S=\int d(vol)\bigg\{\l-{1\over 2 \kappa^2} R+\a_1 R^2 + \a_2 R_{\m\n} R^{\m\n} + \a_3 R_{\m\n\r\s} R^{\m\n\r\s}\bigg\}
\ee
where  the diff invariant measure is given by
\be
d(vol)\equiv \sqrt{|g|} d^n x
\ee
In $n=4$ dimensions the Gauss-Bonnet formula permits to discard  the Riemann squared term; but in higher dimensions this is not so, so we prefer to keep the whole set of operators. Let us analyze them in turn.
In the appendix we have collected  the expansion of the different monomials up to quadratic order in the perturbartions, which is enough to compute the one loop divergences.
\section{The scalar sector of quadratic gravity.}
We are to restrict now the graviton fluctuations to the scalar sector as explained in the introduction, i.e $h_{\m\n}=\phi\bg_{\m\n}$

In the case of the $R$ squared action
\be S=\int d^4x\sqrt{-g} R^2\ee
the EoM is
\be -2\bR_{\m\n}\bR+\frac{1}{2}\bg_{\m\n}\bR^2+2\bn_\m\bn_\n\bR-2\bg_{\m\n}\bar{\Box}\bR=0\ee
which trace yields
\be \label{traceR}\frac{n-4}{2}\bR^2-2(n-1)\bar{\Box}\bR=0\ee
with the ansatz $h_{\m\n}=\phi\bg_{\m\n}$, the expansion \eqref{R2} reduces to 
\begin{align}
S^\phi_{R^2}=\int d^nx \sqrt{\bg}\phi\Bigg\{&(n-1)^2\bar{\Box}^2+\frac{\left(-n^2+7n-6\right)}{2}\bR\bar{\Box}+\frac{n^2-10n+24}{8}\bR^2\Bigg\}\phi\end{align}
we can write
\bea 
S^\phi_{R^2}=(n-1)^2\int d^nx \sqrt{\bg}\phi F\phi\eea
where the explicit expression for this operator is then
\bea
F&=\bar{\Box}^2+\bar{D}^{\m\n}\bn_\m\bn_\n+\bar{P}\eea
where
\bea
\bar{D}^{\m\n}&&=\frac{\left(-n^2+7n-6\right)}{2(n-1)^2}\bg^{\m\n}\bR\nonumber\\
\bar{P}&&=\frac{n^2-10n+24}{8(n-1)^2}\bR^2
\eea

The form of the usual four-dimensional one-loop counterterm is tabulated in  \cite{Barvinsky}, namely,
\bea\label{BV}
\Delta S&&=-\frac{1}{(4\pi)^2}\frac{2}{\e}\int d^n x~\sqrt{|\bg|}\text{tr}\Big\{\frac{1}{180}\left(2\bR_{\m\n\r\s}\bR^{\m\n\r\s}-2\bR_{\m\n}\bR^{\m\n}+5\bR^2\right) \mathbb{I}+\nonumber\\
&&+\frac{1}{48}\bar{D}^2+\frac{1}{24}\bar{D}_{\m\n}\bar{D}^{\m\n}+\frac{1}{12}\bar{D}\bR-\frac{1}{6}\bar{D}^{\m\n}\bR_{\m\n}-\bar{P}+\frac{1}{6}W_{\m\n}W^{\m\n}\Big\}\eea
in our case
\bea\Delta S^\phi_{R^2}
&&=-\frac{1}{(4\pi)^2}\frac{2}{\e}\int d^n x \Bigg\{ \frac{(n-1)^2}{90}\bR_{\m\n\r\s}^2-\frac{(n-1)^2}{90}\bR_{\m\n}^2+\nonumber\\
&&+\frac{3n^4-54n^3+196n^2+424n-1424}{576}\bR^2\Bigg\}\nonumber\\\eea
on-shell, using \eqref{traceR}
\bea\Delta S^\phi_{R^2}
&&=-\frac{1}{(4\pi)^2}\frac{2}{\e}\int d^n x \Bigg\{ \frac{(n-1)^2}{90}\bR_{\m\n\r\s}^2-\frac{(n-1)^2}{90}\bR_{\m\n}^2+\nonumber\\
&&+\frac{(3n^4-54n^3+196n^2+424n-1424)(n-1)}{144(n-4)}\bar{\Box}\bR\Bigg\}\nonumber\\\eea

We continue with the next monomial, Ricci squared
\be S=\int d^4x\sqrt{-g} R_{\m\n}^2\ee
the EoM is 
\be -2\bR_{\m\l}\bR^\l_{~\n}+\frac{1}{2}\bg_{\m\n}\bR_{\a\b}^2-\bar{\Box}\bR_{\m\n}+2\bn_\l\bn_\n\bR_{\m}^{~\l}-\bg_{\m\n}\bn_\a\bn_\b\bR^{\a\b}=0\ee
which trace is
\be \label{traceRicci}\frac{n-4}{2}\bR_{\a\b}^2-\bar{\Box}\bR-(n-2)\bn_\a\bn_\b\bR^{\a\b}=0\ee
with the approximation $h_{\m\n}=\phi\bg_{\m\n}$, in this case \eqref{Ricci2} reduces to 
\begin{align}
S^\phi_{R_{\m\n}^2}=\int d^nx \sqrt{\bg}\phi\Bigg\{&\frac{n}{4}(n-1)\bar{\Box}^2+\bR\bar{\Box}+\frac{\left(-n^2+6n-8\right)}{4}\bR^{\m\n}\bn_\m\bn_\n+\nonumber\\
&+\frac{n^2-10n+24}{8}\bR_{\m\n}^2\Bigg\}\phi\end{align}
again, we can write
\bea 
S^\phi_{R_{\m\n}^2}=\frac{n}{4}(n-1)\int d^nx \sqrt{\bg}\phi F\phi\eea
where the explicit expression for this operator is then
\bea
F&&=\bar{\Box}^2+\bar{D}^{\m\n}\bn_\m\bn_\n+\bar{P}\eea
where
\bea
\bar{D}^{\m\n}&&=\frac{1}{n(n-1)}\Big[4\bg^{\m\n}\bR+\left(-n^2+6n-8\right)\bR^{\m\n}\Big]\nonumber\\
\bar{P}&&=\frac{n^2-10n+24}{2n(n-1)}\bR_{\m\n}^2
\eea
using the one-loop counterterm \eqref{BV}, we obtain
\bea\Delta S^\phi_{R_{\m\n}^2}
&&=-\frac{1}{(4\pi)^2}\frac{2}{\e}\int d^n x\Bigg\{ \frac{n(n-1)}{360}\bR_{\m\n\r\s}^2+\frac{-109n^4+1388n^3-4504n^2+2400n+960}{1440n(n-1)}\bR_{\m\n}^2+\nonumber\\&&+\frac{5n^4-64n^3+8n^2-96n+192}{576n(1-n)}\bR^2\Bigg\}\nonumber\\\eea
on-shell, using \eqref{traceRicci}
\bea\Delta S^\phi_{R_{\m\n}^2}
&&=-\frac{1}{(4\pi)^2}\frac{2}{\e}\int d^n x\Bigg\{ \frac{n(n-1)}{360}\bR_{\m\n\r\s}^2+\frac{-109n^4+1388n^3-4504n^2+2400n+960}{720n(n-1)(n-4)}\bar{\Box}\bR+\nonumber\\&&+\frac{(-109n^4+1388n^3-4504n^2+2400n+960)(n-2)}{720n(n-1)}\bn_\m\bn_\n\bR^{\m\n}+\nonumber\\
&&+\frac{5n^4-64n^3+8n^2-96n+192}{576n(1-n)}\bR^2\Bigg\}\nonumber\\\eea

And finally, the action for Riemann squared
\be S=\int d^4x\sqrt{-g} R_{\m\n\r\s}^2\ee
the EoM is 
\be -2\bR_{\m\a\b\l}\bR_\n^{~\a\b\l}+\frac{1}{2}\bg_{\m\n}\bR_{\a\b\r\s}^2-4\bn^\r\bn^\s\bR_{\m\r\n\s}=0\ee
which trace is 
\be\label{traceRiemann}\frac{n-4}{2}\bR_{\a\b\r\s}^2-4\bn^\r\bn^\s\bR_{\r\s}=0\ee
with approximation $h_{\m\n}=\phi\bg_{\m\n}$, in this case \eqref{Riemann2} reduces to 
\begin{align}
S^\phi_{R_{\m\n\r\s}^2}=\int d^nx \sqrt{\bg}\phi\Bigg\{&(n-1)\bar{\Box}^2+\bR\bar{\Box}+(4-n)\bR^{\m\n}\bn_\m\bn_\n+\frac{n^2-10n+24}{8}\bR_{\m\n\r\s}^2\Bigg\}\phi\end{align}
one time more, we can write
\bea 
S^\phi_{R_{\m\n\r\s}^2}=(n-1)\int d^nx \sqrt{\bg}\phi F\phi\eea
the explicit expression for this operator is then
\bea
F&=\bar{\Box}^2+\bar{D}^{\m\n}\bn_\m\bn_\n+\bar{P}\eea
where
\bea
\bar{D}^{\m\n}&&=\frac{1}{(n-1)}\Big[\bg^{\m\n}\bR+(4-n)\bR^{\m\n}\Big]\nonumber\\
\bar{P}&&=\frac{n^2-10n+24}{8(n-1)}\bR_{\m\n\r\s}^2
\eea
and like the previous case, with \eqref{BV}, we obtain
\bea\Delta S^\phi_{R_{\m\n\r\s}^2}
&&=-\frac{1}{(4\pi)^2}\frac{2}{\e}\int d^n x\Bigg\{ \frac{-45n^2+454n-1084}{360}\bR_{\m\n\r\s}^2+\frac{71n^2-412n+476}{360(n-1)}\bR_{\m\n}^2+\nonumber\\&&+\frac{2n^2+5n+38}{72(n-1)}\bR^2\Bigg\}\eea
on-shell, using \eqref{traceRiemann}
\bea\Delta S^\phi_{R_{\m\n\r\s}^2}
&&=-\frac{1}{(4\pi)^2}\frac{2}{\e}\int d^n x\Bigg\{ \frac{-45n^2+454n-1084}{45(n-4)}\bn^\r\bn^\s\bR_{\r\s}+\frac{71n^2-412n+476}{360(n-1)}\bR_{\m\n}^2+\nonumber\\&&+\frac{2n^2+5n+38}{72(n-1)}\bR^2\Bigg\}\eea

This results shows clearly that the truncation made on the quantum fluctuations is not self-consistent; it is known \cite{Stelle} that the full model is renormalizable; so that there will be mixing between different spins at the quantum level. We pal to study this in the future.

\subsection{Four dimensional check.}
Note that in the  physical dimension $n=4$ 
\be S^\phi_{R_{\m\n\r\s}^2}=S^\phi_{R_{\m\n}^2}=\frac{1}{3}S^\phi_{R^2}=\int d^nx \sqrt{\bg}\phi\left[3\bar{\Box}^2+\bR\bar{\Box}\right]\phi\ee
obviously
\be S^\phi_{R^2}-4S^\phi_{R_{\m\n}^2}+S^\phi_{R_{\m\n\r\s}^2}=0\ee
and the same with the counterterms
\be \Delta S^\phi_{R_{\m\n\r\s}^2}=\Delta S^\phi_{R_{\m\n}^2}=\frac{1}{3}\Delta S^\phi_{R^2}=-\frac{1}{(4\pi)^2}\frac{2}{\e}\int d^n x \Bigg\{\frac{1}{30} \bR_{\m\n\r\s}^2-\frac{1}{30} \bR_{\m\n}^2+\frac{5}{12} \bR^2\Bigg\}\ee
again
\be \Delta S^\phi_{R^2}-4\Delta S^\phi_{R_{\m\n}^2}+\Delta S^\phi_{R_{\m\n\r\s}^2}=0\ee
which is consistent with the Gauss-Bonnet theorem.
\section{Simple scalar model.}
As we have just seen, the scalar sector of quadratic gravity reduces to a scalar lagrangian of a particular type. Let us examine  a particular scalar model which is  closely related to it, namely
\be \label{Ds} S=\int d^nx \phi\left[\bar{\Box}^2+(M^2-m^2)\bar{\Box}-M^2m^2\right]\phi\ee
we can write
\bea 
S_2=\int d^nx \phi F\phi\eea
the explicit expression for this operator is then
\be
F=\bar{\Box}^2+\bar{D}^{\m\n}\bn_\m\bn_\n+\bar{P}\ee
where
\bea
\bar{D}^{\m\n}&&=\bg^{\m\n}(M^2-m^2)\nonumber\\
\bar{P}&&=-M^2m^2
\eea
the one-loop counterterm, with \eqref{BV}, reads
\bea\Delta S_{Mm}
=-\frac{1}{(4\pi)^2}\frac{2}{\e}\int d^n x&& \Big[\frac{1}{180}\left(2\bR_{\m\n\r\s}^2-2\bR_{\m\n}^2+5\bR^2\right)+\frac{(n-2)}{12}(M^2-m^2)\bR+\nonumber\\
&&+\frac{n(n+2)}{48}(M^2-m^2)^2+M^2m^2\Big]\eea
but the same action can be written like
\be S=\int d^nx \sqrt{\bg}\phi\left[-\bar{\Box}-M^2\right]\left[-\bar{\Box}+m^2\right]\phi\ee
the counterterm of operator $-\Box-M^2$ is 
\be\Delta S_{M}
=-\frac{1}{(4\pi)^2}\frac{2}{\e}\int d^n x \Big[\frac{1}{360}\left(2\bR_{\m\n\r\s}^2-2\bR_{\m\n}^2+5\bR^2\right)+\frac{1}{6}M^2\bR+\frac{1}{2}M^4\Big]\nonumber\\\ee
and the corresponding to $-\Box+m^2$ is
\be\Delta S_{m}
=-\frac{1}{(4\pi)^2}\frac{2}{\e}\int d^n x \Big[\frac{1}{360}\left(2\bR_{\m\n\r\s}^2-2\bR_{\m\n}^2+5\bR^2\right)-\frac{1}{6}m^2\bR+\frac{1}{2}m^4\Big]\nonumber\\\ee
if and only if $n=4$ the {\em multiplicative anomaly} \cite{Kontsevich} cancels
\be \Delta S_{Mm}=\Delta S_M+\Delta S_m\ee
This result is interesting insofar as the vanishing of the {\em product anomaly} in this case  seems at variance with a general theorem in \cite{Elizalde}. Let us elaborate. The product anomaly is conventionally defined \cite{Kontsevich} as
\be
a_C(A,B)\equiv \log\det\,(AB)-\log\,\det\,A-\log\det\,B
\ee
and it is known to be non vanishing in general when determinants are defined through the zeta functions associated to the corresponding operators \cite{Ray} and to be given by the Wodzicki residue \cite{JMGB}. 
Our result indicates that in our case the anomaly vanishes
\be
a_C=0
\ee

 In fact, in \cite{Elizalde} the anomaly $a_C$ was computed for the operators
\bea
A\equiv -\Box+m^2_1\nonumber\\
B\equiv -\Box+m^2_2
\eea
in the case $m_2=0$, with the result in terms of the digamma function
\be
a_C(A,B)= {V_n\over (4\pi)^{n/2}}{(-1)^{n/2}\over 2 \left({n\over 2}\right)!} \,m_1^n\left(\Psi(1)-\Psi(n/2)\right)\neq 0
\ee
We do not understand the reason for this discrepancy.

\hrulefill

\section{Some comments on unitarity of the effective action.}
\bi
\item 

In perturbation theory the Kallen-Lehmann spectral theorem \cite{Schwartz} (for a scalar theory, to simplify things)
asserts that the {\em exact} propagator in Minkowski space can be expressed as
\be
\left\langle \Omega \left|T\phi(x)\phi(y)\right|\Omega\right\rangle=\int{d^4 p\over (2\pi)^4} e^{i p(x-y)}\,\int_0^\infty {\rho(s)\over p^2-s+i\e}
\ee
where the {\em spectral function} $\r(s) \geq 0$. In fact the free Feynman propagator 
\be
\Delta(p)=\int_0^\infty {\rho(s)\over p^2-s+i\e}
\ee
can be easily obtained out of the euclidean one through a Wick rotation
\be
\Delta_E(p)\equiv {-1\over p_e^2+m^2}\longrightarrow \Delta_F(p)\equiv {1\over p^2-m^2+i\e}
\ee
This sheds light on what is wrong with ghosts and/or tachyons.  Ghosts have got the wrong sign for residues at the pole; tachyons instead have the pole located at spacelike momenta.
\par
We would like to get a similar clear understanding from the heat kernel point of view.
Let us concentrate in situation at the origin of spatial coordinates, id est, the point $x=y=z=0$. The well defined one-dimensional heat kernel 
\be
K(\t)\equiv \tr\,e^{-\cO \t}
\ee
corresponding to the positive operator
\be
{\cO}\equiv-{d^2\over dt^2}+m^2
\ee
with plane wave eigenfunctions
\be
e^{i\omega t}
\ee
reads, properly normalized,
\be
K(\t)={1\over \sqrt{4\pi\t}}\,e^{-{t^2\over 4 \t}-m^2\t}
\ee
Now it is clear what happens when we want to compute with $-\Box$ (this is just the simplest ghost). Then we should start with

\be
K(\t)\equiv \tr\,e^{+\cO \t}
\ee
and this does not exist. There is a solution of the heat equation, namely
\be
K_{gh}(\t)={1\over \sqrt{4\pi\t}}\,e^{+{t^2\over 4 \t}-m^2\t}
\ee
which however does not reduce to a Dirac's delta when $\t\rightarrow 0$.  
\be
\lim_{\e\rightarrow 0^+}\,{1\over \e \sqrt{\pi}}\,e^{-{x^2\over \e^2}}=\d(x)\neq \lim_{\e\rightarrow 0^+}\,{1\over \e \sqrt{\pi}}\,e^{x^2\over \e^2}
\ee
\par
This is then an ultraviolet problem.
\par
Something similar happens when $m^2 <0$ (the simplest tachyon).
\be
K_{tq}\equiv {1\over \sqrt{4\pi\t}}\,e^{-{t^2\over 4 \t}+m^2\t}
\ee
 The problem there does not lie in the non existence of the small proper time limit, but rather is that the propagator does not exist, because the integral 

\be
\int_0^\infty d\t\,K_{tq}(\t)
\ee 
does not converge. This is an infrared problem.

\item
Let us now comment on Wick's rotation.
The path integral  is defined by
\be
 e^{iW}=\int \mathcal{D}\phi e^{iS}
 \ee
Assume a massive scalar in physical Minkowski space
\be
S\equiv {1\over 2}\int dx_0\,d^3x\,\phi\left(-\Box-m^2\right) \phi
\ee
Perform now the analytic continuation to Euclidean space. It is determined by the need for the mass term to be negative definite, in order for the path integral to formally converge. This implies
\be
x_0=-i x_4
\ee
It follows that the relationship between the differential operators is simply 
\be
\Box\rightarrow-\Box_E
\ee
 which is negative definite. This means that
\be
S\longrightarrow -{i\over 2}\int dx_4 d^3x\,\phi\left(\Box_E-m^2\right)\phi
\ee
The Euclidean effective action is then proportional to
\be
\int dx_4 dx\,\log\,\det\, {\cal O}_E\rightarrow i\int dx_0 dx\,\log\,\det\, {\cal O}_E\equiv i S_{eff}
\ee
in conclusion, this means that the Minkowskian effective action is related to the euclidean determinant through
\be
S_{eff}=\int dx_0 dx\,\log\,\det\, {\cal O}_E
\ee
so that an imaginary piece in the euclidean determinant means an imaginary part in the Minkowskian effective action.

 \item 
 The $\zeta$-function can be recovered from the heat kernel through
\bea
\zeta_{KG}(s)&&={1\over \Gamma(s)}\int_0^\infty {d\t\over \t^{1-s}} \tr\, K(\t;x,y)={1\over \Gamma(s)}\int_0^\infty {d\t\over \t^{1-s}}{1\over (4\pi\t)^{n\over 2}} e^{-m^2\t}=\nonumber\\
&&={\left(m^2\right)^{{n\over 2}-s}\over (4\pi)^{n\over 2}}\,\frac{\Gamma\left(s-{n\over 2}\right)}{\Gamma(s)}
\eea
in even dimension, $n\in 2 \mathbb{N}$,
\be
\frac{\Gamma\left(s-{n\over 2}\right)}{\Gamma(s)}=\frac{1}{(s-1)(s-2)\ldots \left(s-{n\over 2}\right)}
\ee
so that
\be
\zeta_{KG}(s)={\left(m^2\right)^{{n\over 2}-s}\over  (4\pi)^{n\over 2}(s-1)(s-2)\ldots \left(s-{n\over 2}\right)}
\ee
it follows that
\be
\zeta_{KG}(0)= {m^{n+2}\over  (4\pi)^{n\over 2} (-1)^{n\over 2} \left({n\over 2}\right)!}
\ee

In the tachyonic case the proper time integral clearly diverges. We can define it by analytic continuation from the real Klein-Gordon case
\be
\zeta_{tachyon}(s)={\left(-m^2\right)^{{n\over 2}-s}\over  (4\pi)^{n\over 2}(s-1)(s-2)\ldots \left(s-{n\over 2}\right)}
\ee 
that is, we  would get  an extra factor of
\be
\left(-1\right)^{{n\over 2}-s}=e^{i{\pi\over 2}({n\over 2}-s)}
\ee
which yields an imaginary part in the effective action.

\ei

\subsection{ The heat kernel of  the d'Alembertian squared.}
Let is return to  the model \eqref{Ds}
\be S=\int d^nx \phi\left[\bar{\Box}^2+(M^2-m^2)\bar{\Box}-M^2m^2\right]\phi\ee
and consider it as defined in flat space.
The characteristic equation reads
\be
k^4+(m^2-M^2)k^2-M^2 m^2=0
\ee
whose solutions are
\be
k^2= {M^2-m^2\pm\sqrt{(M^2-m^2)^2+4 M^2 m^2}\over 2}=\begin{pmatrix}M^2\\-m^2\end{pmatrix}
\ee
This means that there are runaway classical solutions as long as $m^2\neq 0$. It is to be expected that this carries in a ghostly form to the quantum regime.
\par

Let us try to obtain an exact solution to the heat equation at least in flat space.
\be{\pd \over \pd \t} K(\t,x)=\left[\bar{\Box}^2+(M^2-m^2)\bar{\Box}-M^2m^2\right] K(\t,x)\ee
we can take a Fourier transform
\be K(\t,x)=\frac{1}{(2\pi)^{n+1}}\int F(\omega,k) e^{i \omega \t + i \vec{k}\cdot \vec{x}} d^{n} k d\omega\ee
this implies
\be F(\omega,k)=f(k)\d (-i\omega+k^4+(M^2-m^2)k^2+M^2m^2)\ee
and 
\be K(\t,x)=\frac{1}{(2\pi)^{n}}\int e^{\left[k^4+(M^2-m^2) k^2-M^2m^2\right]\t+i \vec{k}\cdot \vec{x}} f(k)\ee
the boundary condition at $\t=0$ does imply
\be f(k)=1\ee
then we can calculate the integral
\bea K(\t,x)&&= \frac{1}{(2\pi)^{n}}\int d^n k e^{\left[k^4+(M^2-m^2) k^2-M^2m^2\right]\t+i \vec{k}\cdot \vec{x}}=\nonumber\\
&&=\frac{1}{(2\pi)^{n}}\frac{\pi^{\frac{n-1}{2}}}{\Gamma\left(\frac{n+1}{2}\right)}\int_{-1}^1 d\mu\int_0^\infty d k k^{n-1}e^{\left[k^4+(M^2-m^2) k^2-M^2m^2\right]\t+i kx\mu}=\nonumber\\
&&=\frac{1}{(2\pi)^{n}x}\frac{2\pi^{\frac{n-1}{2}}}{\Gamma\left(\frac{n+1}{2}\right)}\int_0^\infty d k k^{n-2}e^{\left[k^4+(M^2-m^2) k^2-M^2m^2\right]\t}\sin{kx}\eea

First of all, it is clear that the integral does not converge unless the coefficient of $|k|^4$ is negative, which we will assume from now on. Then, in the particular case $M=m$ there is an explicit solution
\bea K(\t,x)&&=\frac{1}{(2\pi)^{n}}\frac{\pi^{\frac{n-1}{2}}}{\Gamma\left(\frac{n+1}{2}\right)}\frac{\t^{-\frac{n+2}{4}}}{12}e^{-M^4\t}\Bigg[-x^2\Gamma\left(\frac{n+2}{4}\right){}_pF_q\left[\left\{\frac{n+2}{4}\right\},\left\{\frac{5}{4},\frac{3}{2},\frac{7}{4}\right\},\frac{x^4}{256\t}\right]+\nonumber\\
&&+6\sqrt{\t}\Gamma\left(\frac{n}{4}\right){}_pF_q\left[\left\{\frac{n}{4}\right\},\left\{\frac{1}{2},\frac{3}{4},\frac{5}{4}\right\},\frac{x^4}{256\t}\right]\Bigg]\eea
where the modified hypergeometric function is defines as
\be
\,_pF_q\left(\left\{a_1\ldots a_p\right\},\left\{b_1\ldots b_q\right\},z\right)\equiv \sum{\left(a_1\right)_k\ldots \left(a_p\right)_k\over \left(b_1\right)_k\ldots \left(b_q\right)_k}\,{z^k\over k!}
\ee
and the Pochhammer's symbol is defined as

\be
\left(a\right)_n\equiv{\Gamma(a+n)\over \Gamma(a)}
\ee

\hrulefill

Let us come back to the issue of the sign of the $\Box^2$ term (or the $\Box$ one for that matter). In \cite{Hawkingz} the scaling of determinants when the operator in question is multiplied by a constant factor $\l$ has been studied. The corresponding zeta function reads
\be \zeta_\l(s)=\sum_{n}(\l\l_n)^{-s}=\l^{-s}\zeta(s)\ee
It follows that
\be {d\zeta_\l(s)\over d s} =-\l^{-s}\zeta(s)\log\l+\l^{-s}\zeta'(s)
\ee
For the determinant itself
\be \log\det \Delta\equiv -\zeta'(0)\ee
In the basic case when  $\Delta=\Box$
\be 
\log\det(\l\Box)=\zeta(0)\log\l+\log\det(\Box)
\ee
so that when  $\l=-1$
\be 
\label{minusbox}\log\det(-\Box)=i\pi\zeta(0)+\log\det(\Box)
\ee
which carries over an imaginary part in the effective action, clearly demonstrating violation of unitarity.

\section{Conclusions}
In this paper we have discussed the divergent structure of the scalar spin 0 sector of the graviton fluctuations in theories of gravitation quadratic in curvature. All those contributions have a similar structure, which we summarized in a simplified scalar model, which we studied in some detail.
\par
Although the general lagrangian is known to be renormalizable, owing to Stelle's \cite{Stelle} pioneering work, there is mixing between different spins, and the scalar sector is not. It shares however, many of the properties of the general lagrangian, like runaway solutions and the related presence of ghosts.
\par

We were finally interested in the manifestation of non-unitarity in our language. It is a nontrivial problem, because the heat kernel is {\em defined} as an analytic continuation from the riemannian solution. In the analysis of the scalar model we have found that the {\em product anomaly} vanishes in this particular case, which seems to challenge some theorems on the contrary. 
\par
We were able, nevertheless to discover some imaginary contributions to the effective action for ghosts, although the quantum manifestation of the classical runaway solutions still eludes us in the general case. This is still work in progress.
\section{Acknowledgements}
 We acknowledge partial financial support by the Spanish MINECO through the Centro de excelencia Severo Ochoa Program  under Grant CEX2020-001007-S  funded by MCIN/AEI/10.13039/501100011033
 All authors acknowledge the European Union's Horizon 2020 research and innovation programme under the Marie Sklodowska-Curie grant agreement No 860881-HIDDeN and also byGrant PID2019-108892RB-I00 funded by MCIN/AEI/ 10.13039/501100011033 and by ``ERDF A way of making Europe'' 

\appendix
\section{Expansion of the general action to quadratic order in quantum fluctuations.}
In this section we present the expansion of the different monomials up to quadratic order in the perturbations
\be g_{\m\n}=\bg_{\m\n}+h_{\m\n}
\ee
Let us examine the different monomials in order.
First, we begin with the $R$ squared action
\be S=\int d^4x\sqrt{-g} R^2\ee
the action up to quadratic order in the perturbations yields
\begin{align}\label{R2}
S_{R^2}=\int d^nx \sqrt{\bg}h^{\m\n}\Big\{&\bn_\m\bn_\n\bn_\r\bn_\s-\bg_{\m\n}\bar{\Box}\bn_\r\bn_\s-\bg_{\r\s}\bn_\m\bn_\n\bar{\Box}+\bg_{\m\n}\bg_{\r\s}\bar{\Box}^2+\nonumber\\
&+\frac{\bR}{2}\Big[\bg_{\m\n}\bn_\r\bn_\s+\bg_{\r\s}\bn_\m\bn_\n+\d_{\m\r,\n\s}\bar{\Box}-\bg_{\m\n}\bg_{\r\s}\bar{\Box}-\nonumber\\
&-\frac{1}{4}\left(\bg_{\n\s}\bn_\m\bn_\r+\bg_{\m\s}\bn_\n\bn_\r+\bg_{\n\r}\bn_\m\bn_\s+\bg_{\m\r}\bn_\n\bn_\s\right)-\nonumber\\
&-\frac{1}{4}\left(\bg_{\n\s}\bn_\r\bn_\m+\bg_{\m\s}\bn_\r\bn_\n+\bg_{\n\r}\bn_\s\bn_\m+\bg_{\m\r}\bn_\s\bn_\n\right)\Big]+\nonumber\\
&+\bR_{\m\n}\left(\bg_{\r\s}\bar{\Box}-\bn_\r\bn_\s\right)+\bR_{\r\s}\left(\bg_{\m\n}\bar{\Box}-\bn_\m\bn_\n\right)-\frac{\bR^2}{4}\left(\d_{\m\r,\n\s}-\frac{1}{2}\bg_{\m\n}\bg_{\r\s}\right)+\nonumber\\
&+\frac{\bR}{2}\Big[\frac{3}{4}\left(\bg_{\n\s}\bR_{\m\r}+\bg_{\m\s}\bR_{\n\r}+\bg_{\n\r}\bR_{\m\s}+\bg_{\m\r}\bR_{\n\s}\right)-\bg_{\m\n}\bR_{\r\s}-\bg_{\r\s}\bR_{\m\n}-\nonumber\\
&-\frac{1}{2}\left(\bR_{\m\s\r\n}+\bR_{\n\s\r\m}\right)\Big]+\bR_{\m\n}\bR_{\r\s}\Big\}h^{\r\s}\end{align}
\par
Let us snow turn to the Ricci squared monomial
\be S=\int d^4x\sqrt{-g} R_{\m\n}^2\ee
the action up to quadratic order in the perturbations is
\begin{align}\label{Ricci2}
S_{R_{\m\n}^2}=\int d^nx \sqrt{\bg}h^{\m\n}\Big\{&\frac{1}{2}\bn_\m\bn_\n\bn_\r\bn_\s-\frac{1}{4}\bg_{\m\n}\bar{\Box}\bn_\r\bn_\s-\frac{1}{4}\bg_{\r\s}\bn_\m\bn_\n\bar{\Box}-\nonumber\\
&-\frac{1}{16}\left(\bg_{\m\r}\bn_\n\bn_\s\bar{\Box}+\bg_{\n\r}\bn_\m\bn_\s\bar{\Box}+\bg_{\m\s}\bn_\n\bn_\r\bar{\Box}+\bg_{\n\s}\bn_\m\bn_\r\bar{\Box}\right)-\nonumber\\
&-\frac{1}{16}\left(\bg_{\m\r}\bn_\s\bn_\n\bar{\Box}+\bg_{\n\r}\bn_\s\bn_\m\bar{\Box}+\bg_{\m\s}\bn_\r\bn_\n\bar{\Box}+\bg_{\n\s}\bn_\r\bn_\m\bar{\Box}\right)-\nonumber\\
&+\frac{1}{4}\bg_{\m\n}\bg_{\r\s}\bar{\Box}^2+\frac{1}{4}\d_{\m\r,\n\s}\bar{\Box}^2+\nonumber\\
&+\frac{1}{16}\left(\bg_{\m\s}\bR_{\n\r}\bar{\Box}+\bg_{\n\s}\bR_{\m\r}\bar{\Box}+\bg_{\m\r}\bR_{\n\s}\bar{\Box}+\bg_{\n\r}\bR_{\m\s}\bar{\Box}\right)+\nonumber\\
&+\frac{1}{16}\left(\bR_{\n\s}\bn_\m\bn_\r+\bR_{\m\s}\bn_\n\bn_\r+\bR_{\n\r}\bn_\m\bn_\s+\bR_{\m\r}\bn_\n\bn_\s\right)+\nonumber\\
&+\frac{1}{16}\left(\bR_{\n\s}\bn_\r\bn_\m+\bR_{\m\s}\bn_\r\bn_\n+\bR_{\n\r}\bn_\s\bn_\m+\bR_{\m\r}\bn_\s\bn_\n\right)+\nonumber\\
&+\frac{3}{8}\left(\bR_{\m\r\n\s}\bar{\Box}+\bR_{\n\r\m\s}\bar{\Box}\right)+\frac{1}{2}\d_{\m\r,\n\s}\bR_{\l\t}\bn^\l\bn^\t+\nonumber\\
&+\frac{1}{8}\left(\bg_{\r\s}\bR_{\m\l}\bn_\n\bn^\l+\bg_{\r\s}\bR_{\n\l}\bn_\m\bn^\l+\bg_{\m\n}\bR_{\r\l}\bn_\s\bn^\l+\bg_{\m\n}\bR_{\s\l}\bn_\r\bn^\l\right)+\nonumber\\
&+\frac{1}{8}\left(\bg_{\r\s}\bR_{\m\l}\bn^\l\bn_\n+\bg_{\r\s}\bR_{\n\l}\bn^\l\bn_\m+\bg_{\m\n}\bR_{\r\l}\bn^\l\bn_\s+\bg_{\m\n}\bR_{\s\l}\bn^\l\bn_\r\right)-\nonumber\\
&-\frac{3}{32}\left(\bg_{\n\s}\bR_{\m\l}\bn_\r\bn^\l+\bg_{\m\s}\bR_{\n\l}\bn_\r\bn^\l+\bg_{\n\r}\bR_{\m\l}\bn_\s\bn^\l+\bg_{\m\r}\bR_{\n\l}\bn_\s\bn^\l+\right.\nonumber\\
&\left.+\bg_{\s\n}\bR_{\r\l}\bn_\m\bn^\l+\bg_{\s\m}\bR_{\r\l}\bn_\n\bn^\l+\bg_{\r\n}\bR_{\s\l}\bn_\m\bn^\l+\bg_{\r\m}\bR_{\s\l}\bn_\n\bn^\l\right)-\nonumber\\
&-\frac{3}{32}\left(\bg_{\n\s}\bR_{\m\l}\bn^\l\bn_\r+\bg_{\m\s}\bR_{\n\l}\bn^\l\bn_\r+\bg_{\n\r}\bR_{\m\l}\bn^\l\bn_\s+\bg_{\m\r}\bR_{\n\l}\bn^\l\bn_\s+\right.\nonumber\\
&\left.+\bg_{\s\n}\bR_{\r\l}\bn^\l\bn_\m+\bg_{\s\m}\bR_{\r\l}\bn^\l\bn_\n+\bg_{\r\n}\bR_{\s\l}\bn^\l\bn_\m+\bg_{\r\m}\bR_{\s\l}\bn^\l\bn_\n\right)-\nonumber\\
&-\frac{1}{8}\left(\bR_{\m\l\n\s}\bn_\r\bn^\l+\bR_{\n\l\m\s}\bn_\r\bn^\l+\bR_{\m\l\n\r}\bn_\s\bn^\l+\bR_{\n\l\m\r}\bn_\s\bn^\l\right)-\nonumber\\
&-\frac{1}{8}\left(\bR_{\m\l\n\s}\bn^\l\bn_\r+\bR_{\n\l\m\s}\bn^\l\bn_\r+\bR_{\m\l\n\r}\bn^\l\bn_\s+\bR_{\n\l\m\r}\bn^\l\bn_\s\right)-\nonumber\\
&-\frac{1}{4}\bg_{\m\n}\bg_{\r\s}\bR_{\l\t}\bn^\l\bn^\t-\frac{1}{4}\d_{\m\r,\n\s}\bR_{\a\b}^2+\frac{1}{8}\bg_{\m\n}\bg_{\r\s}\bR_{\a\b}^2-\nonumber\\
&-\frac{1}{2}\left(\bg_{\m\n}\bR_{\r}^{~\l}\bR_{\s\l}+\bg_{\r\s}\bR_{\m}^{~\l}\bR_{\n\l}\right)+\frac{1}{8}\left(\bR_{\m\r}\bR_{\n\s}+\bR_{\n\r}\bR_{\m\s}\right)+\nonumber\\
&+\frac{5}{16}\left(\bg_{\n\s}\bR_{\m}^{~\l}\bR_{\r\l}+\bg_{\m\s}\bR_{\n}^{~\l}\bR_{\r\l}+\bg_{\n\r}\bR_{\m}^{~\l}\bR_{\s\l}+\bg_{\m\r}\bR_{\n}^{~\l}\bR_{\s\l}\right)-\nonumber\\
&-\frac{1}{4}\left(\bR_{\m}^{~\l}\bR_{\n\r\s\l}+\bR_{\n}^{~\l}\bR_{\m\r\s\l}+\bR_{\m}^{~\l}\bR_{\n\s\r\l}+\bR_{\n}^{~\l}\bR_{\m\s\r\l}\right)-\nonumber\\
&-\frac{1}{8}\left(\bR_{\m\l\n\t}\bR^{\t~~\l}_{~\s\r}+\bR_{\n\l\m\t}\bR^{\t~~\l}_{~\s\r}+\bR_{\m\l\n\t}\bR^{\t~~\l}_{~\r\s}+\bR_{\n\l\m\t}\bR^{\t~~\l}_{~\r\s}\right)\Big\}h^{\r\s}\end{align}

And finally we take the Riemann squared action
\be S=\int d^4x\sqrt{-g} R^2_{\m\n\a\b}\ee
the action up to quadratic order in the perturbations reads
\begin{align}\label{Riemann2}
S_{R_{\m\n\r\s}^2}=\int d^nx \sqrt{\bg}h^{\m\n}\Big\{&\bn_\m\bn_\n\bn_\r\bn_\s+\d_{\m\r,\n\s}\bar{\Box}^2-\nonumber\\
&-\frac{1}{4}\left(\bg_{\m\r}\bn_\n\bn_\s\bar{\Box}+\bg_{\n\r}\bn_\m\bn_\s\bar{\Box}+\bg_{\m\s}\bn_\n\bn_\r\bar{\Box}+\bg_{\n\s}\bn_\m\bn_\r\bar{\Box}\right)-\nonumber\\
&-\frac{1}{4}\left(\bg_{\m\r}\bn_\s\bn_\n\bar{\Box}+\bg_{\n\r}\bn_\s\bn_\m\bar{\Box}+\bg_{\m\s}\bn_\r\bn_\n\bar{\Box}+\bg_{\n\s}\bn_\r\bn_\m\bar{\Box}\right)+\nonumber\\
&+\left(\bR_{\m\r\n\s}\bar{\Box}+\bR_{\n\r\m\s}\bar{\Box}\right)+\nonumber\\
&+\frac{1}{8}\Big[-3\left(\bg_{\m\s}\bR_{\n\t\r\l}+\bg_{\n\s}\bR_{\m\t\r\l}+\bg_{\m\r}\bR_{\n\t\s\l}+\bg_{\n\r}\bR_{\m\t\s\l}\right)+\nonumber\\
&+5\left(\bg_{\m\s}\bR_{\n\l\r\t}+\bg_{\n\s}\bR_{\m\l\r\t}+\bg_{\m\r}\bR_{\n\l\s\t}+\bg_{\n\r}\bR_{\m\l\s\t}\right)-\nonumber\\
&-\left(\bg_{\m\s}\bR_{\n\r\l\t}+\bg_{\n\s}\bR_{\m\r\l\t}+\bg_{\m\r}\bR_{\n\s\l\t}+\bg_{\n\r}\bR_{\m\s\l\t}\right)\Big]\left(\bn^\l\bn^\t+\bn^\t\bn^\l\right)-\nonumber\\
&-\frac{1}{4}\left(\bg_{\m\n}\bR_{\r\l\s\t}+\bg_{\m\n}\bR_{\s\l\r\t}+\bg_{\r\s}\bR_{\n\l\m\t}+\bg_{\r\s}\bR_{\m\l\n\t}\right)\left(\bn^\l\bn^\t+\bn^\t\bn^\l\right)+\nonumber\\
&+\d_{\m\r,\n\s}\bR_{\l\t}\bn^\l\bn^\t-\nonumber\\
&-\frac{1}{8}\left(\bg_{\n\s}\bR_{\m\l}\bn^\l\bn_\r+\bg_{\m\s}\bR_{\n\l}\bn^\l\bn_\r+\bg_{\n\r}\bR_{\m\l}\bn^\l\bn_\s+\bg_{\m\r}\bR_{\n\l}\bn^\l\bn_\s+\right.\nonumber\\
&\left.+\bg_{\s\n}\bR_{\r\l}\bn^\l\bn_\m+\bg_{\s\m}\bR_{\r\l}\bn^\l\bn_\n+\bg_{\r\n}\bR_{\s\l}\bn^\l\bn_\m+\bg_{\r\m}\bR_{\s\l}\bn^\l\bn_\n\right)-\nonumber\\
&-\frac{1}{8}\left(\bg_{\n\s}\bR_{\m\l}\bn^\l\bn_\r+\bg_{\m\s}\bR_{\n\l}\bn^\l\bn_\r+\bg_{\n\r}\bR_{\m\l}\bn^\l\bn_\s+\bg_{\m\r}\bR_{\n\l}\bn^\l\bn_\s+\right.\nonumber\\
&\left.+\bg_{\s\n}\bR_{\r\l}\bn^\l\bn_\m+\bg_{\s\m}\bR_{\r\l}\bn^\l\bn_\n+\bg_{\r\n}\bR_{\s\l}\bn^\l\bn_\m+\bg_{\r\m}\bR_{\s\l}\bn^\l\bn_\n\right)+\nonumber\\
&+\frac{1}{2}\left(\bR_{\n\s}\bn_\m\bn_\r+\bR_{\m\s}\bn_\n\bn_\r+\bR_{\n\r}\bn_\m\bn_\s+\bR_{\m\r}\bn_\n\bn_\s\right)+\nonumber\\
&+\frac{1}{2}\left(\bR_{\n\s}\bn_\r\bn_\m+\bR_{\m\s}\bn_\r\bn_\n+\bR_{\n\r}\bn_\s\bn_\m+\bR_{\m\r}\bn_\s\bn_\n\right)-\nonumber\\
&-\frac{1}{4}\left(\bg_{\m\s}\bR_{\n\r}\bar{\Box}+\bg_{\n\s}\bR_{\m\r}\bar{\Box}+\bg_{\m\r}\bR_{\n\s}\bar{\Box}+\bg_{\n\r}\bR_{\m\s}\bar{\Box}\right)+\nonumber\\
&+\left(\frac{1}{8}\bg_{\m\n}\bg_{\r\s}-\frac{1}{4}\d_{\m\r,\n\s}\right)\bR^2_{\a\b\l\t}+\nonumber
\end{align}
\raggedbottom
\begin{align}
&+\frac{3}{4}\left(\bg_{\m\s}\bR_{\n\a\b\l}\bR_\r^{~\a\b\l}+\bg_{\n\s}\bR_{\m\a\b\l}\bR_\r^{~\a\b\l}+\bg_{\m\r}\bR_{\n\a\b\l}\bR_\s^{~\a\b\l}+\bg_{\n\r}\bR_{\m\a\b\l}\bR_\s^{~\a\b\l}\right)-\nonumber\\
&-\frac{1}{2}\left(\bg_{\m\n}\bR_{\r\a\b\l}\bR_\s^{~\a\b\l}+\bg_{\r\s}\bR_{\m\a\b\l}\bR_\n^{~\a\b\l}\right)+\nonumber\\
&+\frac{1}{4}\left(\bR_{\m\l}\bR^\l_{~\s\n\r}+\bR_{\n\l}\bR^\l_{~\s\m\r}+\bR_{\m\l}\bR^\l_{~\r\n\s}+\bR_{\n\l}\bR^\l_{~\r\m\s}\right)+\nonumber\\
&+\frac{1}{4}\left(\bR_{\m\r\l\t}+\bR_{\m\l\r\t}+\bR_{\r\m\l\t}+\bR_{\r\l\m\t}\right)\bR^{\l~~\t}_{~\s\n}+\nonumber\\
&+\frac{1}{4}\left(\bR_{\n\s\l\t}+\bR_{\n\l\s\t}+\bR_{\s\n\l\t}+\bR_{\s\l\n\t}\right)\bR^{\l~~\t}_{~\r\m}-\nonumber\\
&-\frac{1}{4}\left(\bg_{\n\r}\bR_{\m\l}\bR_\s^{~\l}+\bg_{\m\r}\bR_{\n\l}\bR_\s^{~\l}+\bg_{\n\s}\bR_{\m\l}\bR_\r^{~\l}+\bg_{\m\s}\bR_{\n\l}\bR_\r^{~\l}\right)\Big\}h^{\r\s}\end{align}



\begin{thebibliography}{99}
 \bibitem{AASG}
E.~Alvarez, J.~Anero and S.~Gonzalez-Martin,
``Quadratic gravity in first order formalism,''
JCAP \textbf{10} (2017), 008
doi:10.1088/1475-7516/2017/10/008
[arXiv:1703.07993 [hep-th]].\\
E.~Alvarez, J.~Anero and R.~Santos-Garcia,
``Structural stability of spherical horizons,''
[arXiv:2006.02463 [hep-th]].\\
``Weyl anomalies and the nature of the gravitational field,''
[arXiv:1907.03781 [hep-th]].\\
``Weighing the Vacuum Energy,''
Phys. Rev. D \textbf{103}, no.8, 084032 (2021)
doi:10.1103/PhysRevD.103.084032
[arXiv:2011.08231 [hep-th]].
\bibitem{Alvarez:2016}
E.~Alvarez and S.~Gonzalez-Martin,
``Weyl Gravity Revisited,''
JCAP \textbf{02} (2017), 011
doi:10.1088/1475-7516/2017/02/011
[arXiv:1610.03539 [hep-th]].\\
\bibitem{Ray}
D.~B.~Ray and I.~M.~Singer,
``Analytic torsion for complex manifolds,''
Annals Math. \textbf{98} (1973), 154-177
doi:10.2307/1970909
\bibitem{Kontsevich}
M.~Kontsevich and S.~Vishik,
``Geometry of determinants of elliptic operators,''
[arXiv:hep-th/9406140 [hep-th]].

\bibitem{Elizalde}
E.~Elizalde, L.~Vanzo and S.~Zerbini,
``Zeta function regularization, the multiplicative anomaly and the Wodzicki residue,''
Commun. Math. Phys. \textbf{194} (1998), 613-630
doi:10.1007/s002200050371
[arXiv:hep-th/9701060 [hep-th]].

\bibitem{JMGB}
J.~M.~Gracia-Bondia, J.~C.~Varilly and H.~Figueroa,
``Elements of noncommutative geometry,''




\bibitem{Windows}
E.~Alvarez,
``Windows on Quantum Gravity,''
Fortsch. Phys. \textbf{69} (2021) no.1, 2000080
doi:10.1002/prop.202000080
[arXiv:2005.09466 [hep-th]].

\bibitem{Schwartz}
``Quantum field theory and the standard model"
(Cambridge,2015)
 \bibitem{Alvarez:2018lrg}
 E.~Alvarez, J.~Anero, S.~Gonzalez-Martin and R.~Santos-Garcia,
 ``Physical content of Quadratic Gravity,''
 Eur. Phys. J. C \textbf{78} (2018) no.10, 794
 doi:10.1140/epjc/s10052-018-6250-x
 [arXiv:1802.05922 [hep-th]].
 \bibitem{Barnes}
 K. J. Barnes, Unpublished (part of Ph. D. Thesis at University of London) (1963)
 R. J. Rivers, ``Lagrangian Theory or Neutral Massive Spin-2 Field'', Nuov. Cim. 34
 (1964) 386.
	\bibitem{Barvinsky} 
 A.~O.~Barvinsky and G.~A.~Vilkovisky,
 ``The Generalized Schwinger-Dewitt Technique in Gauge Theories and Quantum Gravity,''
 Phys.\ Rept.\  { 119}, 1 (1985).
 doi:10.1016/0370-1573(85)90148-6

\bibitem{Buchbinder}
I.~L.~Buchbinder, S.~D.~Odintsov and I.~L.~Shapiro,
``Effective action in quantum gravity,''\\
I.~L.~Buchbinder and I.~Shapiro,
``Introduction to Quantum Field Theory with Applications to Quantum Gravity,''

\bibitem{Dirac}
P.~A.~M.~Dirac,
``Classical theory of radiating electrons,''
Proc. Roy. Soc. Lond. A \textbf{167} (1938), 148-169
doi:10.1098/rspa.1938.0124

\bibitem{Einhorn}
M.~B.~Einhorn and D.~R.~T.~Jones,
``Renormalizable, asymptotically free gravity without ghosts or tachyons,''
Phys. Rev. D \textbf{96} (2017) no.12, 124025
doi:10.1103/PhysRevD.96.124025
[arXiv:1710.03795 [hep-th]].\\
P.~G.~Ferreira, C.~T.~Hill, J.~Noller and G.~G.~Ross,
``$R^2$/Higgs inflation and the hierarchy problem,''
[arXiv:2108.06095 [hep-ph]].\\
A.~Salvio and A.~Strumia,
``Agravity,''
JHEP \textbf{06} (2014), 080
doi:10.1007/JHEP06(2014)080
[arXiv:1403.4226 [hep-ph]].
E as of 02 Feb 2022
\bibitem{Grinstein}
B.~Grinstein, D.~O'Connell and M.~B.~Wise,
``The Lee-Wick standard model,''
Phys. Rev. D \textbf{77}, 025012 (2008)
doi:10.1103/PhysRevD.77.025012
[arXiv:0704.1845 [hep-ph]].
 \bibitem{Hawkingz}
 Hawking, S.W. ``Zeta function regularization of path integrals in curved spacetime.'' Commun.Math. Phys. 55, 133–148 (1977). https://doi.org/10.1007/BF01626516
\bibitem{Hawking}
S.~W.~Hawking and T.~Hertog,
``Living with ghosts,''
Phys. Rev. D \textbf{65} (2002), 103515
doi:10.1103/PhysRevD.65.103515
[arXiv:hep-th/0107088 [hep-th]].

\bibitem{Kontsevich}
M.~Kontsevich and S.~Vishik,
``Geometry of determinants of elliptic operators,''
[arXiv:hep-th/9406140 [hep-th]].
\bibitem{Stelle}
K.~S.~Stelle,
``Renormalization of Higher Derivative Quantum Gravity,''
Phys. Rev. D \textbf{16} (1977), 953-969
doi:10.1103/PhysRevD.16.953\\
``Classical Gravity with Higher Derivatives,''
Gen. Rel. Grav. \textbf{9} (1978), 353-371
doi:10.1007/BF00760427

 \bibitem{York}
 J.~W.~York, Jr.,
 ``Conformally invariant orthogonal decomposition of symmetric tensors on Riemannian manifolds and the initial value problem of general relativity,''
 J. Math. Phys. \textbf{14} (1973), 456-464
 doi:10.1063/1.1666338
 

\end{thebibliography}
\end{document}